# Stellar Coronagraphy

**Eugene Serabyn[1] and Michael Bottom[2]**

[1]Jet Propulsion Laboratory, California Institute of Technology, Pasadena, CA 91125, USA

[2]Institute for Astronomy, University of Hawaii at Manoa, 640 N. Aohoku Pl, Hilo, HI 96720, USA

**Abstract** Detecting exoplanets and other faint sources of emitted and reflected light near a bright star requires deeply suppressing the starlight while efficiently transmitting the dim light from its surroundings. This suppression can be carried out by coronagraphs, nulling interferometers, and starshades. This chapter provides a brief overview of these technologies, emphasizing coronagraphs.

## Introduction

Stars are typically surrounded by numerous smaller bodies and diffuse emission sources, ranging from stellar coronae to dust, asteroids, comets, exoplanets, moons and planetary rings, with the central stars outshining all of these at most wavelengths. Indeed, due to the smaller sizes, lower temperatures and/or diffuseness of these secondary sources, the flux ratio, or "contrast", between such off-axis emission sources and the host star can be very small (e.g., Seager and Deming 2010). Moreover, all such off-axis sources lie very near their host star from our vantage point, making them exceedingly difficult to pick out of the bright halo of starlight that typically surrounds stellar images (e.g., Fig. 1). Suppression of starlight and its attendant photon noise is therefore essential to being able to detect faint planets and other circumstellar emission. This issue was first addressed in the context of the Sun by Lyot (1939), who devised the first coronagraph to block the solar disk to image its corona. Modern solar coronagraphs are described in Rabin (2021).

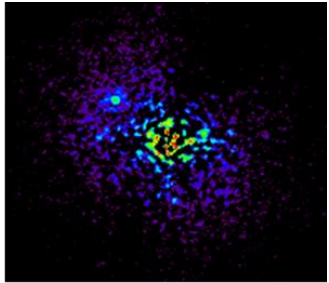

Fig. 1. Image of κ And acquired with a vortex coronagraph on the well-corrected subaperture of the Palomar Hale Telescope. κ And B is at upper left.

On the other hand, the goal of imaging planets and circumstellar dust around nearby stars (exoplanets and exozodiacal dust, respectively) has more recently led to the development of a number of stellar suppression techniques, including stellar coronagraphs (Guyon et al. 2006, Galicher & Mazoyer 2023) that suppress starlight after its arrival at a telescope, interferometric "nullers" that destructively interfere the starlight collected by multiple telescopes or telescope sub-apertures (Beichman et al. 1999, Serabyn 2021), and "starshades," spacecraft that block starlight before it can reach a distant telescope (Spitzer 1962, Cash 2006). The ultimate goals of exoplanet observations are comparative planetology and the spectroscopy of terrestrial, i.e., Earth-like, exoplanets (Astro2020), but more modest contrast cases, such as young, thermally-emitting Jovian planets and circumstellar dust disks, are already accessible with current ground- and space-based coronagraphs. This chapter briefly summarizes the optical techniques involved, focusing mainly on coronagraphy, which is already in use is space, but other promising techniques are also included. For brevity, all starlight suppression techniques will be referred to generically in the following as coronagraphy.

## Exoplanet Detection Requirements

Stellar coronagraphy is challenging because the light reflected by an exoplanet is many orders of magnitude fainter than the direct starlight. The fraction of the emitted starlight subtended by the planet is $(r/a)^2/4$, where $r$ is the planet's radius and $a$ its distance from the star. Incorporating also a wavelength-dependent albedo (reflectivity) and time-dependent orbital phase function (the fraction of the illuminated planetary surface visible to the observer, analogous to the phases of the moon), yields a contrast of $A(\lambda)\Phi(t)(r/a)^2/4$. For Jupiter-like and Earth-like planets at optical wavelengths, the contrast relative to the Sun is then $\approx 10^{-9}$ and $10^{-10}$, respectively (e.g., Seager & Demming 2010).

Directly detecting an exoplanet requires separating its light from its host star's. Unfortunately, light, including starlight, cannot be focused to an arbitrarily small point. Diffraction due to finite-sized telescope apertures will broaden stellar images into the form of the Airy pattern (Born & Wolf 2011), with relatively bright concentric "Airy rings" that often lie at angular separations comparable to planet-star separations for nearby stars. Moreover, wavefront aberrations due either to imperfect telescope and instrument optics or to atmospheric turbulence above ground-based telescopes will lead to further image degradation. Both the diffraction pattern due to the telescope and scattered light due to wavefront errors must be suppressed, for which coronagraphy and wavefront correction, respectively, are necessary.

Before discussing specific instrumental implementations, it is important to consider what a stellar suppression system must do. First, an angular resolution fine enough to separate faint close-in exoplanets from their host stars is required, which calls for large telescope diameters. Specifically, at distances of 10 – 20 pc from us, an Earth-analog on a 1 Astronomical Unit radius orbit would appear at a maximum separation from its host star of 50 - 100 milli-arc seconds (mas). On the other hand, the half-power diffraction beamwidth of a telescope of diameter D is $\approx \lambda/D$, where $\lambda$ is the wavelength of observation, i.e., $\approx 200\ \lambda_{\mu m}/D_m$ mas, where $\lambda_{\mu m}$ is the wavelength in microns and $D_m$ the diameter in meters. With the diffraction beamwidth proportional to wavelength, the worst resolution will be at the longest observing wavelength, which for silicon-based detectors is at $\lambda \approx 1\ \mu m$. One must also consider the "inner working angle" (IWA), i.e., the "closest" to a star that an exoplanet can be detected with a given technique, as all stellar suppression systems also strongly attenuate exoplanet light very near the star. The IWA is typically taken to be the half-power point of a system's radial transmission profile, but there is no sharp detection cutoff at the IWA. For an IWA of, e.g., two diffraction beamwidths, (i.e., 2 $\lambda/D$), the ability to resolve exoplanets from host stars within 10 to 20 pc, at wavelengths up to 1 $\mu m$, then calls for minimum telescope diameters of $\approx 4 – 8$ m.

Flux levels are also an important consideration, as an exoplanet with $10^{-10}$ contrast relative to a 5th magnitude (m = 5) solar-like star would have m = 30, or $\approx 0.01$ phot/m$^2$/sec over a typical astronomical filter of 10% bandwidth. Ideally, starlight would be suppressed sufficiently to reduce the stellar photon noise below that due to other astrophysical noise sources, such as the photon noise due to the exoplanet itself and to the zodiacal and exozodiacal light present in the field of view (Stark et al. 2014). This requires very accurate wavefront correction to remove scattered starlight in a dark image-plane search-space termed the "dark hole" (Malbet et al. 1995). An example is shown in Fig. 2. Most importantly, the exoplanet photon detection rate must allow spectroscopy in reasonable integration times. This again calls for a large telescope to maximize collecting area. Finally, for the noise to be dominated by astrophysical noise sources, very low-noise detectors are required. The integration times for an ~6 m telescope with a coronagraph to measure spectra of nearby terrestrial exoplanets lie in the range of a few weeks to a few months, depending on the particular star (Stark et al. 2014, Stark et al. 2019, Morgan et al. 2023, Zhang et al. 2024, Mennesson et al. 2024), highlighting the critical importance of large telescopes, high optical throughputs, and low-noise detectors.

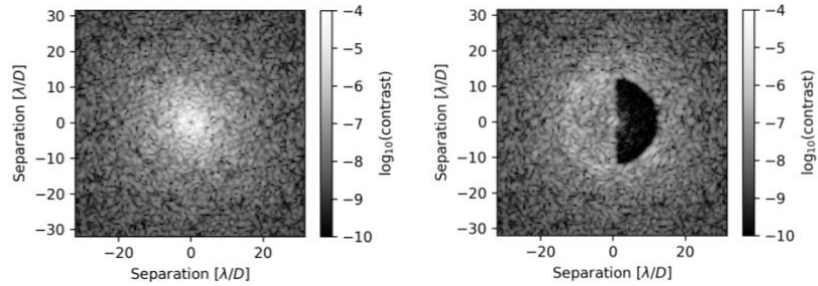

Fig. 2. Left: Speckled stellar image after an ideal coronagraph. Right: The image after wavefront correction with a single DM, resulting in a one-sided dark hole. Both images simulated with HCIPy (Por et al. 2018).

Of course, comparative planetology will require stellar rejection across a wide wavelength range, potentially from the near-ultraviolet (0.2 μm) to the near-infrared (1.8 μm), in order to access the numerous tracers of atmospheric properties and potential biomarkers (e.g., Astro2020). With both individual coronagraphic masks and wavefront correction solutions likely to apply only over limited wavelength ranges, several coronagraphic observations in different sub-bands will likely be necessary to cover the full spectral region desired, each of which will require long integrations, unless done in parallel.

Beyond these requirements on angular resolution, photometric sensitivity and wavelength coverage, the main requirement specific to starlight rejection is that the essentially on-axis starlight must be rejected to a high degree while the nearby exoplanet light is efficiently transmitted. A starlight suppression system is thus essentially a spatial filter with a sharp transition from low to high transmission at some small IWA. This rapid transition is normally aided in practice by the generation of the off-axis dark-hole by means of wavefront control, as in Fig. 2. The magnitude of the suppression needed can be appreciated by considering the brightness of the inner stellar Airy rings in the absence of a coronagraph – specifically, the second Airy ring (at 2.7 $\lambda/D$) has a peak brightness relative to the central stellar peak of $4 \times 10^{-3}$ (Born & Wolf 2011), so seeing a terrestrial exoplanet at that radius requires a reduction in starlight there by a factor $> 10^7$.

## Wavefront Quality Considerations

The separation of the exoplanet and starlight signals cannot be carried out effectively unless the stellar point spread function (PSF) is close to ideal. However, passage down a real optical beam train or through the Earth's atmosphere invariably leads to aberrated wavefronts, making adaptive optics wavefront correction essential to high-contrast imaging. How good must the wavefront be? While a full analysis would consider the wavefront-error power spectrum, a simple estimate can be made by considering the action of an N × N element deformable mirror (DM) on the root mean square (rms) wavefront phase error, $\varphi_{rms}$ (with $\varphi$ in radians). For small $\varphi_{rms}$, the total fraction of scattered starlight is given by $\varphi_{rms}^2 = 1 - S$, where S is the Strehl ratio (Born & Wolf 2011). If this light were scattered only over a dark hole of size N × N, the contrast, c, would be

$$c \sim \frac{\varphi_{rms}^2}{N^2},$$

so the phase accuracy needed to reach a given contrast level is

$$\varphi_{rms} \sim N\sqrt{c}.$$

A $10^{-10}$ contrast with, e.g., a 100 × 100 element DM then implies $\varphi_{rms} \sim 10^{-3}$, i.e., ~ $10^{-4}$ wavelengths, or 50 pm for $\lambda$ = 500 nm. To first order, each of the DM actuators must therefore be set to a height accuracy finer than the diameter of a hydrogen atom. Finally, note that the dark hole's maximum extent, set by the Nyquist limit to the wavefront correction that can be applied with an N × N element DM, is $\frac{N\lambda}{2D}$ in "radius". However, its extent is often more limited, especially as a single DM cannot correct both wavefront phase and amplitude on both sides of a point source.

Of course, wavefront correction to such a level requires wavefront sensing to even better levels. According to the uncertainty principle, measurement of a wave's phase to an accuracy of $\Delta\phi$ is accompanied by a photon number uncertainty, $\Delta n$, of $\Delta n > \frac{1}{2\Delta\phi}$ (Loudon 2000). Assuming Poisson statistics, roughly $10^6$ photons are then needed per wavefront measurement per DM segment area, which gives the stellar flux needed. With each DM actuator controlling a portion of the primary mirror of size, e.g., 10 cm, and, assuming a 10% throughput to the coronagraphic focal plane (assuming wavefront sensing in the focal plane), the needed photon count is $10^9$ photons/m$^2$ within the measurement time and passband. Nearby main sequence stars, the primary targets for imaging searches, can provide these photon counts in ≈ 1 min, assuming several (≈ 4) intensity measurements required to measure phase (Wyant 1975). However, to detect exoplanets at $10^{-10}$ contrast at several sigma significance, the noise must be correspondingly lower. Also allowing for an iterative approach to the final wavefront solution on a given star, wavefront correction timescales can stretch into hours, thus calling for an ultra-stable telescope and optical beam train (Stahl et al. 2020). On the other hand, ground-based coronagraphs, which must correct atmospheric fluctuations on millisecond timescales, are necessarily limited by the wavefront sensing constraint to poorer contrasts by a few orders of magnitude (Guyon 2018).

Any wavefront phase measurement ultimately relies on intensity measurements (e.g., Gerchberg and Saxton 1972), as the oscillation frequency of a light wave in the optical or infrared is many orders of magnitude faster than any detector response time. The most direct methods rely solely on the science image, i.e., without the use of a separate wavefront sensor at an intermediate point in the beam train. The goal of wavefront correction is then to reduce the residual scattering from wavefront errors -- often referred to as speckles (see Fig. 2) -- surrounding the stellar image. As described in Malbet et al. (1995) and Borde and Traub (2006), every sinusoidal pupil-plane phase-error leads to a symmetrically placed pair of image plane speckles, which can be corrected by applying the opposite wavefront shape to a pupil-plane DM. A number of algorithms for such speckle suppression exist, with the simplest being based on the application of small sinusoidal probe waves of various spatial frequencies and lateral phases to a DM to "null" speckles, and the more complex approaches such as Electric Field Conjugation (EFC) also including a model of the optical system (Borde & Traub 2006, Give'on et al. 2007, Bottom et al. 2016, Guyon 2018, Poitier 2020, 2022, Guyon et al. 2021).

However, at very deep contrasts, wavefront errors can arise due both to phase (i.e., shape) and amplitude (i.e., reflectivity) errors, while a pupil-plane DM only introduces phase shifts. A second DM in a non-pupil plane is therefore needed, with the Talbot effect used to change phase errors into amplitude errors upon propagation (Goodman 2017). A second DM also allows correction in a full 360° field of view around the star (Borde & Traub 2006). Finally, additional high-speed wavefront sensors are also needed to maintain pointing and to control low-order modes prior to starlight removal to guarantee optimal coronagraph performance (Guyon et al. 2009, Shi et al. 2018).

## Coronagraph Types

Coronagraphic starlight suppression techniques all modify the incident stellar light waves prior to detection. The amplitude or phase of the wavefront can be modified to redirect, diffract, block or attenuate the on-axis starlight, while still passing the slightly off-axis exoplanet light. Optical masks to carry out these functions are in an image plane and/or a pupil plane, as illustrated in Fig. 3. The DM may also play a role. The operation of coronagraphic masks can be understood by means of Fourier optics, with the main point being that in the proper optical configuration (e.g., Fig. 3), the electric field distributions in successive focal and pupil planes are Fourier transforms of each other (Goodman 2017). Key to any coronagraphic approach is the set of optical masks used to implement it, meaning that mask designs required for a particular approach must also be manufacturable.

The following sections discuss the four main coronagraph types: amplitude and phase coronagraphs with masks either in the focal or pupil planes. Different coronagraph types have different advantages and disadvantages in terms of performance metrics such as contrast, IWA and throughput, which allows for various performance tradeoffs, but also implies that no single coronagraph is optimal from all points of view. As space does not permit discussing every possible coronagraph here, we highlight didactic cases as well as promising cases that have either already demonstrated very deep contrast in the laboratory or that have been baselined for actual or proposed space missions.

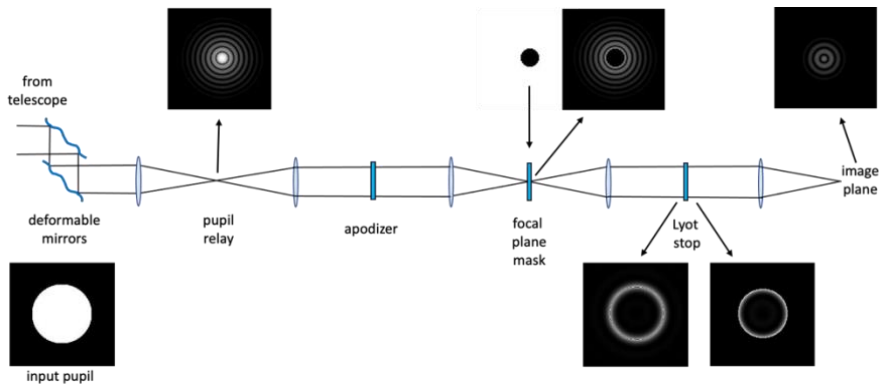

Fig. 3. Schematic coronagraph optical layout, with the Lyot case illustrated. The focal-plane mask (an occulting disk here) lies in an image plane, while one of the DMs, a potential apodizer, and the Lyot stop lie in pupil planes. The three focal-plane images along the top show the input Airy function, the occulted Airy pattern and the final image. Shown along the bottom are the input pupil and the pre- and post-Lyot-stop pupil images. The calculations shown here and in Figs.5 and 8 were carried out with the IRCT coronagraph simulation software of K. Liewer.

## Focal-plane Amplitude Coronagraphs

The goal of a focal-plane "amplitude" coronagraph is to block a fraction of the stellar PSF using an opaque image-plane mask. However, while a small opaque disk-shaped mask can block the center of the stellar Airy pattern, the outer rings pass, as seen in Fig. 3. While larger disks block more starlight, the desire to pass light from exoplanets as near stars as possible limits disk diameters. As the Airy rings arise from the sharp input-pupil edge, their residue will appear in a downstream pupil image concentrated mostly near the pupil rim, as seen in Fig. 3. As suggested by Lyot, a second blocker in the form of an undersized circular pupil-plane stop (the "Lyot stop") is then used to block (part of) this residue. Two distinct masks and rejection steps are thus involved in a Lyot coronagraph: one in the focal plane and a second in a subsequent pupil plane.

The rejection of a classical Lyot coronagraph is affected by the radii of both the opaque disk and the Lyot stop, and these parameters are also coupled, as smaller disk radii diffract residual starlight further into the pupil. Defining a single optimum is thus not possible: a larger disk radius improves starlight rejection but increases the IWA, while a smaller Lyot stop also improves starlight rejection but reduces exoplanet throughput and resolution. A smaller IWA also increases sensitivity to pointing and other low-order wavefront errors, as slightly decentered or aberrated stellar PSFs will more easily leak past a smaller blocker.

The mask's intensity transmission profile need not be a binary on-off function, and numerous alternatives exist, including a Gaussian transmission profile, which partially attenuates more distant Airy rings, and the band-limited profile, which in theory can remove all starlight from a slightly undersized pupil (Kuchner & Traub 2002). However, spatial variations in mask transmission necessarily bring attendant phase shifts, implying the need for (chromatic) phase correction. Using DM-based phase correction, a one-dimensional band-limited mask has reached a contrast of 6 x $10^{-10}$ for a 10% bandwidth of light (Trauger & Traub 2007). Alternatively, the mask itself can include a corrective phase layer (Fig. 4, Trauger et al. 2012, Trauger et al. 2024). Such "hybrid Lyot" masks can be adapted to telescopes with complex (i.e., partially obscured) pupil shapes, and one such mask (Trauger et al. 2016) is to be flown on the Roman Space Telescope's Coronagraph Instrument (CGI).

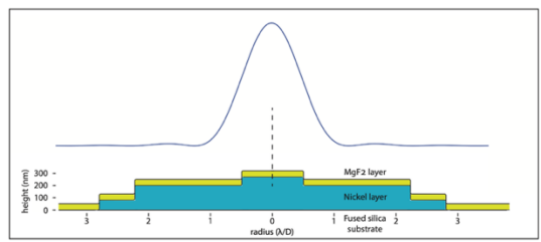

Fig. 4. Cross-section of a hybrid metal-dielectric mask that also includes a central phase dimple for low-order Zernicke wavefront sensing. The curve shows the scale of the focal plane Airy function at 550 nm. Courtesy J. Trauger and NASA (https://exoplanets.nasa.gov/internal_resources/3035).

## Focal-plane Phase Coronagraphs

Alternatively, the phase of the star's focal-plane electric-field distribution can be modified. As a pure phase mask leaves the intensity in that plane unaltered, the role of a focal-plane phase mask is to redirect the starlight, much as phase gratings steer diffracted orders. As a result, starlight blockage in this case occurs entirely at a subsequent pupil-plane Lyot stop. The primary goal of a focal-plane phase mask is then to direct as much starlight as possible outside the pupil, where a Lyot stop can effectively block it with no penalty to planet light.

Many spatial phase distributions can be applied to the focal plane field, with ideal cases leaving zero light interior to the downstream pupil image. Several such solutions exist for the case of a clear (unobscured) pupil. For purely azimuthal phase variations, this includes (at least) two phase families: the even pie-shaped family in which the phase in successive sectors alternates between 0 and π rad, and the even optical vortex family, in which the phase is proportional to the azimuth around the mask center, reaching a phase of $2\pi n$ in a full circuit about the center, where the integer $n$ is called the topological charge of the vortex (Mawet et al. 2005, Foo et al. 2005, Schwartzlander 2009). (The pie-shaped family includes the four-quadrant phase mask [FQPM; Rouan et al. 2000; Fig. 5], and the eight-octant phase mask [8OPM; Murakami et al. 2010], etc.) In both families, the masks with even phase distributions provide zeroed out pupils, while the odd family members, such as the "phase knife" mask (Abe et al. 2001), which consists of two halves at 0 and π rad, as well as vortex masks of odd topological charge, leave significant leakage inside the pupil, as can be seen in these masks' Lyot-plane distributions shown in Fig. 5.

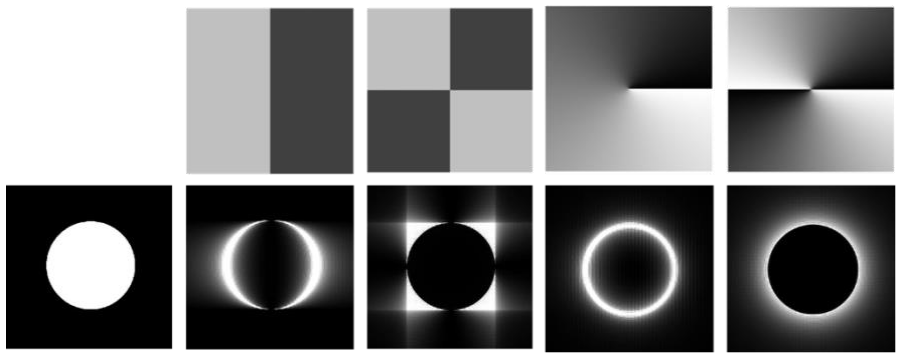

Fig. 5. Top row: Four focal-plane phase masks (left to right: phase knife; FQPM; charge 1 optical vortex; charge 2 optical vortex). The grey scale in these images covers the range 0 (black) to 2π (white) radians (twice for the rightmost image). Bottom row, left to right: the input pupil, and the four downstream pupil-plane images corresponding to the masks above them. Note the complete absence of light within the original pupil for the two even masks (third and fifth panels).

Radial phase variations can also be employed. The simplest case is that of a central circular phase disk that introduces a π radian phase shift (causing destructive interference) between the inner and outer regions of the stellar Airy pattern containing roughly equal power (Roddier & Roddier 1995). However, the main issue facing phase masks, i.e., their chromaticity, impacts phase disks twice, as both the desired thickness of the central layer and its diameter relative to the width of the Airy pattern depend (linearly) on wavelength. Improved broadband performance is possible with an additional phase ring (Fig. 6) around the central region (Soummer et al. 2003).

Purely azimuthal phase masks only have chromaticity due to the layer thicknesses. Considering first longitudinal phase (i.e., phase due to propagation through variable dielectric thicknesses), such chromaticity can be reduced by combining multiple layers of different materials (Swartzlander 2006), but such an approach tends to require individual layer heights too large for typical deposition techniques. Such longitudinal masks are referred to as "scalar" phase masks because longitudinal phase is generally polarization-independent.

On the other hand, use of geometric phase (Pancharatnam 1956) can potentially avoid such chromaticity, by instead using layer birefringence. The desired spatially-variant birefringent layers can be made using, e.g., form birefringence, or liquid crystal polymer (LCP), photonic crystal or metamaterial layers (Mawet et al. 2005, Mawet et al. 2009, Murakami et al. 2013, Palatnick et al. 2023). However, such birefringent, or "vector", masks (which rely on polarization manipulations) have the disadvantage of producing equal and opposite phase ramps in the two circular polarization states, which can lead to ambiguities in wavefront sensing, and may require polarization-splitting to reach very deep contrast. The LCP vector approach has so far reached a contrast of 1.6 x $10^{-9}$ in a 10% bandwidth (Ruane et al. 2022).

As polarization splitting increases instrumental complexity, a high-performance scalar vortex mask would be ideal. Some ideas for less chromatic scalar vortex masks (Fig. 6) include the non-regularly wrapped vortex mask of Galicher et al. (2020), and a mask with both radial and azimuthal phase structures, i.e., a vortex mask with a central phase disk (Desai et al 2024), both of which can improve broadband contrast by roughly a factor of 100.

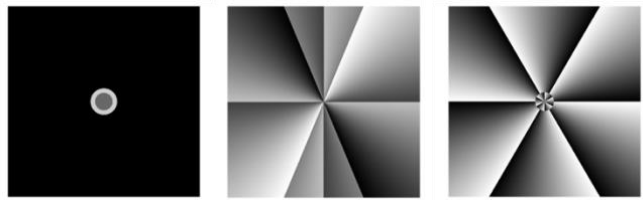

Fig. 6. Several approaches to achromatizing scalar phase masks. Left: The dual-zone phase disk of Soummer et al. 2003). Center: The non-regularly wrapped vortex mask of Galicher et al. (2020). Right: the dimpled charge 6 vortex mask of Desai et al. (2023). The grey scale in all three images goes from black to white for 0 to $2\pi$ radians.

## Pupil-plane Amplitude Coronagraphs

Alternatively, the electric field can be modified in the pupil-plane. The main advantage to altering the pupil plane field is insensitivity to pointing errors, as all angular offsets see the same pupil mask. This is also the main disadvantage, as a planet's PSF will be modified the same way as the star's, eliminating the ability of focal-plane coronagraphs to suppress starlight while allowing planet light through relatively unimpeded.

In pupil-plane amplitude coronagraphs, the pupil amplitude distribution is essentially modified to reshape the focal-plane PSF. According to Fourier theory, a sharply truncated spatial distribution leads to ringing in the complementary spatial-frequency domain (Davis et al. 2001), and such ringing can be reduced by apodizing, i.e., smoothing, the edge of the spatial distribution. In the 2d telescope aperture case, ringing in the spatial-frequency plane is seen as the Airy pattern, and apodization of the aperture results in suppression of the Airy rings.

The simplest apodization to consider is the azimuthally symmetric case, for which an optimal solution is the generalized prolate spheroidal function, which suppresses the Airy rings and concentrates starlight within a few $\lambda/D$ of the center (Slepian 1965, Kasdin et al. 2003). However, apodization has two deleterious effects: the throughput decreases (identically for the stellar and exoplanet light), and the core of the focal-plane PSF is broadened everywhere in the field, degrading the angular resolution. Moreover, the starlight in the reshaped PSF is not actually removed, which may cause saturation and other problems with real detectors, unless the starlight is blocked in a downstream focal plane. Even more problematic is mask manufacturability: smooth spatially-dependent transmission variations are challenging to fabricate, and any spatially-variant attenuator layer will have attendant phase shifts that need to be controlled to the level described earlier.

One can avoid these difficulties by using binary masks, wherein the transmission at any point in the pupil is either 0 or 1 (Kasdin et al. 2005). Here manufacturability is not an issue unless the structures are too fine. As the main design variable is then the shape of the pupil region that transmits, such apodizers are referred to as shaped pupil apodizers. Shaped-pupil apodizers can be tailored to the input aperture shape, and optimized for contrast level, IWA, throughput and dark-hole size targets (Carlotti et al. 2011). It is also easy to hide pupil obstructions such as on-axis secondary mirrors and supports behind zero-transmission regions. Finally, shaped-pupil masks and other types of pupil apodizers can be followed by Lyot-type masks, leading to the Shaped-Pupil Lyot Coronagraph (Kasdin et al. 2020, Eldorado Riggs et al. 2021), a number of which are to be flown in the Roman CGI (Fig. 7), and the apodized pupil Lyot coronagraph (APLC; Peuyo et al. 2017).

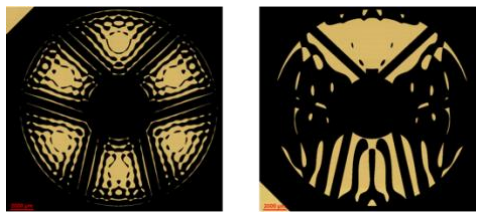

Fig. 7. Two of the binary shaped pupil masks in the Roman CGI, used for wide-field imaging (left) and spectroscopy (right). (Courtesy AJ Eldorado Riggs, J Sandhu and NASA).

**Pupil-plane Phase Coronagraphs**

Pupil-plane phase modification can also be used to shape a PSF, in the same way that pupil-plane wavefront aberrations are known to distort PSFs. Desired phase patterns can be applied either with dedicated phase masks or, for small enough phase shifts, using a DM located in the pupil. A simple example is the application of a comatic wavefront shape to generate an asymmetric PSF with one side darker than the other (e.g., Fig. 8, left panel). This has been demonstrated on sky using a DM (Serabyn et al. 2007), as have more optimized pupil-plane phase distributions (Codona et al. 2006, Kenworthy et al. 2007).

However, with the exoplanet and stellar light subject to identical phase patterns, one cannot overly distort the PSF without also losing exoplanet light from the PSF core. Very good solutions exist, even with apertures vignetted by an on-axis secondary mirror and its supports, such as the phase apodized pupil Lyot coronagraph (Por 2020). In this case, the optimized pupil-plane phase pattern is followed by an opaque knife-edge mask that blocks slightly more than half the focal plane (e.g., Fig. 8), and then a Lyot stop in the downstream pupil. Fig. 8 illustrates these steps for the simple case of a pupil phase given by an exaggerated level of coma. Much deeper one-sided rejection is possible with optimized pupil phases, reaching IWAs as small as 1.4 $\lambda/D$ with reasonably high efficiencies (Por 2020). Of course, somewhat less than half of the focal plane is accessible at any time.

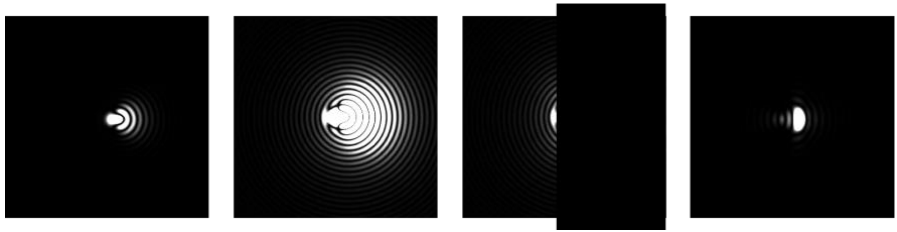

Fig. 8. Left: An asymmetric comatic PSF. The image is saturated, with a range of 0 to $10^{-1}$ of the image peak. Left-center: The same image, but highly saturated to see the outer diffraction rings, with a range of 0 to $10^{-3}$ of the image peak. Right-center: The right-hand side of the last image is blocked to 1.2 $\lambda/D$ left of center with an opaque knife-edge blocker. (On the same scale.) Right: Final image (after the Lyot stop), with a range of 0 to $10^{-4}$ of the original (left hand) image peak.

Pupil-plane phase masks can also be either scalar or vector, with the latter again being polarization-dependent, but potentially easier to make broadband. Moreover, inclusion of phase grating structures within a vector mask provides for the separation of the two circular polarization states on the detector array, with complementary dark holes in the two circular polarization states on opposite sides of the center (Snik et al. 2012, Otten et al. 2014, Doelman et al. 2017, 2021).

PSF shaping can also be brought about by optics that remap the pupil intensity distribution. This "phase-induced amplitude apodization" (PIAA) approach (Guyon 2003, Kern et al. 2013) uses a highly aspheric mirror pair that severely distorts off-axis exoplanet PSFs, usually requiring inversion after starlight removal. Corrected PIAA optics largely conserve flux and resolution, resulting in high throughput in the exoplanet PSF core, but the aspheric mirrors needed are difficult to manufacture.

## Single-mode Nulling Coronagraphy

Most coronagraphs typically suppress diffraction at angles beyond a few $\lambda/D$ of the star (Guyon et al. 2006), but smaller angles can be reached using cross-aperture nulling interferometry, a technique in which anti-phasing parts of the telescope pupil yield an interferometric null, i.e., a dark hole, centered on the stellar position (Fig. 9; Ruane et al. 2018, Echeverri et al. 2019). The rejection is deepened by coupling the central dark region (i.e., the core of the original stellar PSF) to a single-mode fiber or waveguide (Fig. 9), as the anti-symmetric stellar field generated by passage through an appropriate phase mask (or using the DM) cannot couple into a symmetric fiber mode (Haguenauer and Serabyn 2007). However, sources slightly off-axis are transmitted (Fig. 9), which has enabled on-sky companion detections in to $\approx 0.3$ $\lambda/D$ (Serabyn et al. 2019; Echeverri et al. 2024), albeit at modest contrast. Such small IWA techniques may be especially important at longer wavelengths, where stellar diffraction patterns are larger.

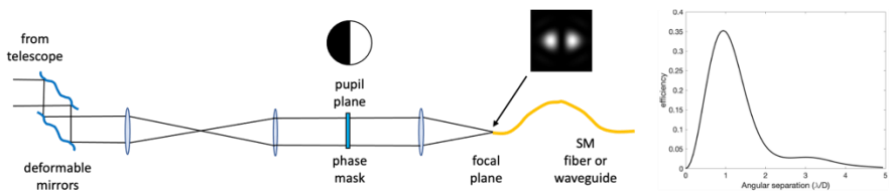

Fig. 9. Left: Single-mode nulling coronagraph layout, with the phase-knife case illustrated. Right: Radial coupling curve calculated using the nulling software packages of K. Liewer and G. Ruane.

The masks needed for single-mode nulling coronagraphy are generally anti-symmetric (to minimize coupling into the symmetric fiber mode), which is the opposite of the even masks needed for coronagraphy, but there is overlap. In Cartesian and polar coordinates, appropriate masks are the phase-knife mask and the optical vortex mask, respectively. Both have yielded laboratory nulls of a few $10^{-5}$ (Echeverri et al. 2019, Serabyn et al. 2024), not yet deep enough for very faint exoplanets. However, more complex pupil-plane masks can provide both deeper and broader stellar nulling, the latter important for partially resolved stars (Serabyn et al. 2024). The potential of single-mode nulling coronagraphy is only beginning to be explored, with further promising areas being the use of integrated optics (Norris et al. 2020) and photonic lanterns (Xin et al. 2022).

**System-Level Implications for Coronagraphs and Nullers**

As coronagraphs and nullers operate behind telescopes, also of great importance are system-level issues. The first is whether the telescope's secondary mirror is on- or off-axis. All coronagraphs work better with a clear unobscured telescope aperture, but some (such as pupil-apodizers) work better than others in the presence of a secondary mirror blockage. Polarization aberrations arise for fast telescope designs, as unequal Fresnel reflection coefficients in the two orthogonal polarization states create two different (and incoherent) PSFs (Chipman, Lam & Breckinridge 2015). Segmented telescopes introduce further undesirable diffraction from segment gaps (Pueyo et al. 2013), though at much lower level than an on-axis secondary mirror and its support struts. Finally, a critical issue for all coronagraphs is the exoplanet light throughput, given not only losses at masks and stops, but also the large number of optics in general. The number of optical elements must thus be minimized, while still enabling coronagraphic operations, including passing the light through the requisite number of focal and pupil plane masks, a final field stop and a spectrometer, as well as allowing tip-tilt and wavefront sensing, wavelength multiplexing, and perhaps even polarization multiplexing.

Of course, one sure way to detect more exoplanet photons is to increase the telescope diameter. However, in addition to raising mission cost, the resultant improvement in angular resolution will increasingly resolve the surfaces of nearby stars, degrading attainable contrast in a manner somewhat similar to pointing errors. Finally, maintaining the dark hole for long periods requires a very stable optical system, including an ultrastable telescope (Stahl et al. 2020). This has numerous implications, including high structural stiffness, a high degree of thermal stabilization, highly accurate wavefront sensing and metrology through the entire optical system to enable measuring slowly varying telescope and panel shapes, an orbit with minimal disturbances (thermal, gravitational…), such as near the L2 Lagrange point, and particle impact protection to avoid the larger impact events seen on JWST. In other words, anything that can degrade the wavefront requires mitigation.

## Stellar Coronagraphs on Space Telescopes

Stellar coronagraphs are currently on board both the Hubble Space Telescope (HST) and the James Webb Space Telescope (JWST). These have provided many beautiful images both of large-separation exoplanets and of bright exozodiacal dust disks (Ren et al. 2023). The HST coronagraphs, including decommissioned ones on the ACS and NICMOS instruments, have largely been simple coronagraphs without any active wavefront control. The remaining HST coronagraph in the Space Telescope Imaging Spectrograph (STIS), consisting of two perpendicular wedged opaque stripes, provides a variable obscuration width in the direction perpendicular to the stripe selected, allowing some flexibility in blocking the PSF and accommodating pointing errors. STIS reaches contrasts of $3 \times 10^{-5}$ at 0.25" and $10^{-6}$ at 0.6" (Debes et al. 2019).

The JWST coronagraphs include both Lyot-type masks and FQPMs, and also lack active wavefront control. Relatively similar contrasts to HST are seen (Girard et al. 2022, Boccaletti et al. 2022), with the larger telescope mirror allowing the resolution of interesting inner dust structures. One unique aspect of JWST coronagraphy is the inclusion of coronagraphic channels well into the mid-infrared, i.e., out to 23 μm, which has provided many interesting new results, including the detection of warm inner exozodiacal dust components and temperate giant planets (Boccaletti et al 2024, Worthen et al. 2024, Matthews et al. 2024).

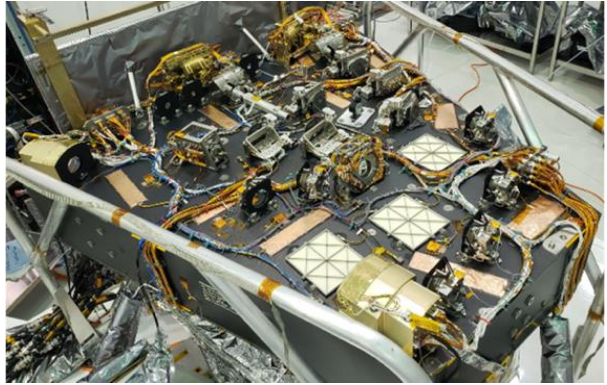

Fig. 10. The Roman CGI instrument. The two DMs are at top left and bottom center (gold color). Courtesy F. Zhao and NASA.

The first space-based coronagraph to include deformable mirrors and active wavefront control will be the Roman Space Telescope's CGI, seen in Fig. 10. With a pair of DMs for wavefront correction, the CGI's contrast at $\lambda = 575$ nm is expected to be better than $10^{-8}$ for separations of 3-9 $\lambda/D$ (Bailey et al. 2023), representing an improvement of $\sim$ 3 orders of magnitude over any other operational coronagraph. The CGI will include both a hybrid Lyot mask and shaped pupil masks. With this contrast, CGI has the potential to provide the first visible-light images of inner exozodiacal disks, and the first reflected-light cold Jupiters (Bailey et al. 2023).

For the more distant future, the recent National Academy of Sciences decadal report (Astro2020) has recommended a space-based coronagraph able to spectroscopically characterize $\approx$ 25 nearby terrestrial exoplanets. In response, NASA is considering a mission concept named the Habitable Worlds Observatory (HWO) aimed at reaching the requisite $10^{-10}$ contrast. The earlier HABEX and LUVOIR mission studies (Gaudi et al. 2019, LUVOIR2020) had each considered "optimal" coronagraphs for their mission concepts, and had selected the vortex coronagraph for the off-axis telescope case, and the APLC for the on-axis secondary case, but the future HWO coronagraphs have not yet been selected, and indeed, no coronagraphic technique has yet demonstrated the desired broadband $10^{-10}$ contrast.

**Starshades**

A conceptually distinct approach to imaging and spectrally characterizing exoplanets is a starshade, a large occulting spacecraft positioned far upstream of a space telescope (Spitzer 1962, Cash 2006). The telescope and starshade align with the target star, blocking the starlight while passing the off-axis planet light. A circular occulting shape would diffract starlight on-axis, into the telescope, so starshades have tapered petals optimized to generate a dark shadow over the full pupil of the telescope (Fig. 11), essentially creating an artificial stellar eclipse (Vanderbei et al. 2007, Cash 2011). Recent experiments to construct a starshade are summarized in Willems et al. (2022).

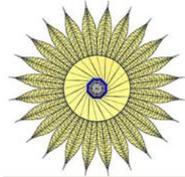

Fig. 11. Illustration of a starshade, from the NASA Exoplanet Program Starshade Technology Development webpage: https://exoplanets.nasa.gov/exep/technology/starshade/

An external starshade has significant advantages and disadvantages relative to an internal coronagraph (Seager et al. 2019, Gaudi et al. 2019). First, starshades can provide small IWAs, the IWA being given approximately by the angle that the starshade's radius subtends. A 50 mas IWA thus implies a starshade distance about four million times its diameter. Second, laboratory demonstrations with small starshade-like occulters have already reached $10^{-10}$ contrasts (Harness et al. 2019). Third, as the starlight is removed before reaching the telescope, neither a high-performance coronagraph nor an ultra-stable telescope is required. The elimination of the coronagraph allows for many times higher optical throughput, important both for reducing observation times and for potentially enabling observations into the ultraviolet, where reflection losses are higher and wavefront control more challenging (Shaklan et al. 2023). Finally, a single starshade can enable a wide bandwidth of operation, with the passband determined by the (adjustable) starshade to telescope separation.

On the other hand, substantial technical and operational challenges exist. The shadow diameter increases with starshade size, and must be larger than the telescope aperture (Glassman et al. 2009). This leads to a minimum starshade diameter of tens of meters to work with a large space telescope, and thus a separation of tens of thousands of kilometers to preserve IWA. The starshade diameter exceeds the fairing size of the largest rockets, so the starshade must be launched in a stowed configuration and unfurled in space, with a deployed shape accuracy of ~ 0.1 mm (McKeithen et al. 2019). The relative alignment must be kept to less than a meter at these large separations to prevent starlight leaking into the telescope, and the position must be actively maintained in the presence of gravity gradients (Bottom et al. 2020, Flinois et al. 2020).

A fundamental operational challenge is the long timescales ($\approx$ weeks) and significant fuel requirements for switching targets, given the wide angular spacing of the bright target stars and the large telescope-starshade separation (Stark et al. 2016). This makes multiple visits to, e.g., determine exoplanet orbits, especially problematic, though multiple starshades can partially address this retargeting issue. Additionally, the time spent reorienting a starshade system can be used by the telescope for other astronomical observations.

The tradeoff is thus between a more sensitive but less flexible multi-spacecraft starshade system, and a less sensitive but more flexible single-telescope coronagraph, with both approaches being quite complex and expensive. While no starshade has yet been flown in space, nor is one planned at present, the starshade's throughput advantage certainly warrants further consideration.

**Summary**

Many starlight suppression concepts have emerged in response to the goal of imaging exoplanets in nearby solar systems. These have resulted in an exploration of a wide range of coronagraph designs, including both amplitude and phase-based approaches, as well as hybrids, making use of electric field manipulations in both focal and pupil planes. At the same time, wavefront control techniques and the DMs available to carry them out have improved, allowing for ever deeper contrasts to be reached in the laboratory. In addition, valuable practical experience is being gained by using coronagraphs on both ground-based and space-based telescopes. Finally, in the near future, the Roman space telescope's CGI will demonstrate coronagraphy to deep enough levels to potentially see cold reflected-light Jupiters and inner exozodiacal disks in visible light, representing a leap in contrast performance of about three orders of magnitude.

However, while recent coronagraphic performance improvements have been impressive, terrestrial exoplanets are both very faint and very close to their host stars, meaning that significant progress is still needed to enable spectroscopic observation of the holy grail: Earth-like exoplanets around Sun-like stars. Indeed, even with the most promising coronagraphic techniques, expected spectroscopic integration times of up to a few months on nearby terrestrial exoplanets mean that astronomers may need to wait for their harvest of photons from such targets as long as farmers must typically wait for their more terrestrial crops.

20